\documentclass[prl,twocolumn,aps,superscriptaddress,showpacs,floatfix]{revtex4}
\usepackage{amsmath,amssymb,latexsym,amsfonts}
\usepackage{epsfig}
\usepackage{colordvi}
\begin{document}
%\title{Wavelength-dependence of high-order harmonic generation}
\title{Quantum path interference in the wavelength dependence of high-harmonic generation}
\date{\today}

\author{K. Schiessl}
\email[Electronic address: ]{klaus@concord.itp.tuwien.ac.at}
\affiliation{Institute for Theoretical Physics, Vienna University of 
Technology, Wiedner Hauptstra\ss e 8-10, A--1040 Vienna, Austria, EU}
\author{K. L. Ishikawa}
\affiliation{Department of Quantum Engineering and Systems Science, 
Graduate School of Engineering, University of Tokyo, Hongo 7-3-1,
Bunkyo-ku, Tokyo 113-8656, Japan\\}

\affiliation{PRESTO (Precursory Research for Embryonic Science and Technology), Japan Science and Technology Agency, Honcho 4-1-8, Kawaguchi-shi, Saitama 332-0012, Japan}
\author{E. Persson}
\affiliation{Institute for Theoretical Physics, Vienna University of 
Technology, Wiedner Hauptstra\ss e 8-10, A--1040 Vienna, Austria, EU}
\author{J. Burgd\"orfer}
\affiliation{Institute for Theoretical Physics, Vienna University of 
Technology, Wiedner Hauptstra\ss e 8-10, A--1040 Vienna, Austria, EU}

\begin{abstract}
\noindent
We investigate the dependence of the intensity of radiation due to high-harmonic generation (HHG) as a function of the wavelength  $\lambda$ of the fundamental driver field.
Superimposed on a smooth power-law dependence observed previously we find surprisingly strong and rapid fluctuations on a fine $\lambda$ scale.
We identify the origin of these fluctuations in terms of quantum path interferences with up to five returning orbits significantly contributing.
\end{abstract}

\maketitle

%\section{Introduction}

\noindent
High harmonic generation (HHG) represents a versatile and highly successful avenue towards an ultrashort coherent
light source  covering a wavelength range from the vacuum ultraviolet to the soft X-ray region \cite{ar:seres05}.
HHG has successfully opened new research areas, such as attosecond science \cite{Hentschel2001Nature,Tsakiris2003Nature} and nonlinear optics in the XUV region \cite{Sekikawa2004Nature,Nabekawa2005PRL}.
The fundamental wavelength $\lambda$ used in most of existing HHG experiments is in the near-visible range ($\sim 800$ nm). 
The cutoff law $E_c=|E_i|+3.17U_p$, where $|E_i|$ denotes the binding energy of the target atom and $U_p=F_0^2/4\omega^2$ the ponderomotive energy ($F_0$: laser electric field strength), suggests that a longer fundamental wavelength is advantageous to extend the cutoff to a higher photon energy, since $U_p$ quadratically increases with $\lambda$. 
There is an increasing interest in the development of high-power mid-infrared ($\sim 2 \mu$m) laser systems, e.g., based on optical parametric chirped pulse amplification. 
Along those lines the dependence of the HHG yield on $\lambda$ has become an issue of major interest.
It has been commonly accepted that the spreading of the returning wavepacket would result in a $\lambda^{-3}$ dependence of the HHG efficiency  \cite{ar:lewenstein94} as long as ground state depletion can be neglected \cite{Gordon2005OE}.
Experimental findings \cite{Shan2001PRA} have provided partial support. 
Recently, however, Tate \emph{et al.} \cite{ar:tate_scaling07} have reported a different wavelength-scaling of HHG between 800 nm and 2 $\mu$m calculated with the time-dependent Schr\"odinger equation (TDSE) for Ar and a strong-field approximation (SFA) for He. 
They found a more rapidly decreasing HHG yield $\propto \lambda^{-x}$  with $5 \le x \le 6$.
This surprising finding based on a somewhat limited number of data points motivated us to explore the $\lambda$ dependence in more detail employing two completely independent integration methods of the TDSE to check for consistency and convergence.
We have investigated the HHG for H and Ar on the level of single-atom response.
Surprisingly, the  harmonic yield does not smoothly decrease with fundamental wavelength but exhibits rapid oscillation with a period of $6-20$ nm depending on the wavelength region. 
A semiclassical analysis based on the SFA reveals that the rapid oscillations are due to the interference of up to five different rescattering trajectories. 
Remarkably, averaged over the fast oscillations the smoothed yield follows an approximate  $\lambda^{-5}$ scaling, qualitatively consistent with the results of Tate \textit{et al.} \cite{ar:tate_scaling07}.

%\section{Numerical Method}
We solve the atomic time-dependent Schr\"{o}dinger equation in a linearly polarized laser field in the length gauge,
\begin{eqnarray}
i\frac{\partial}{\partial t} \psi({\bf r},t) & = & 
\left[-\frac{1}{2}\nabla^2+V_{\rm eff}(r) + z \, F(t) \right] \psi({\bf r},t),
\label{eq:tdse}
\end{eqnarray}
where $F(t)=F_0 f(t) \sin(\omega t)$ denotes the laser electric field, $f(t)$ is the envelope function and $V_{\rm eff}(r)$ the atomic potential.
For H, $V_{\rm eff}(r)$ is the bare Coulomb potential while for Ar we employ a model potential \cite{ar:muller98} within the single-active electron approximation which reproduces the binding energy to an accuracy of typically $ \approx 10^{-3}$. %\Delta E_i
We employ two complementary methods to solve Eq.\ (\ref{eq:tdse}) in order to establish reliable and consistent results.

%\section{Method 1} 
In the first method, Eq.\ (\ref{eq:tdse}) is numerically integrated using the alternating direction
implicit (Peaceman-Rachford) method \cite{Kulander1992} with a uniform grid spacing $\Delta r$ of $6.25\times 10^{-2}$ a.u.
In order to reduce the difference between the discretized and analytical wave function, we scale the
Coulomb potential by a few percent at the first grid point \cite{Krause1992}.
The time step $\Delta t$ is 1/16000 of an optical cycle for 800 nm wavelength, i.e., $6.895\times 10^{-3} {\rm a.u.}$. 
This algorithm is accurate to the order of $\mathcal{O}(\Delta t^3)$. 
%
%\section{Method 2} 
In the second method,  the TDSE is integrated on a finite grid by means of the pseudo-spectral method \cite{ar:tong97} which is also accurate to the order of $\mathcal{O}(\Delta t^3)$. 
It allows for timesteps of the order of 0.1 atomic units.
The $r$ coordinate is discretized within the interval $[0,r_{\rm max}]$ with a non-uniform mesh point distribution. 
The innermost grid point is typically as small as $2.5\times 10^{-4}$ a.u., enabling an accurate description near the nucleus.
A smooth cut-off function is multiplied at each time-step to avoid spurious reflections at the border $r_{\rm max}$, 
while equivalently another cut-off function prevents reflections at the largest resolved energy  $E_{\rm max}$.
Deeply bound, occupied states supported by the model potential are dynamically blocked during the time evolution \cite{ar:klaus_hhg06}.
We calculate the dipole acceleration $\ddot d(t) = -  \partial_t^2 \langle z(t) \rangle $, 
employing the Ehrenfest theorem through the relation $\ddot d(t)= \langle \psi({\bf r},t)\mid \cos \theta /r^2  - F(t) \mid \psi({\bf r},t)\rangle$\cite{ar:tong97}, 
in which the second term can be dropped as it does not contribute to the HHG spectrum.

%\section{Global dependence}
For a direct comparison we adopt the laser parameters of with Ref.\ \cite{ar:tate_scaling07}, with a fixed peak intensity of $1.6 \times 10^{14} $ W/cm$^2$, a variation of $\lambda$ between 800 nm and 2 $\mu$m, 
and an envelope function $f(t) $ corresponding to a 8-cycle flat-top sine pulse with a half-cycle turn-on and turn-off. 
We have checked that the fluctuations in the harmonic yield to be discussed below are not an artefact of this particular choice of $f(t)$. 
They can be observed also for ``smoother'' pulse shape such as a $\sin^2$ pulse, provided that the pulse length is large enough to enable multiple returning trajectories. %$T_p$ 

\begin{figure}
\epsfig{file=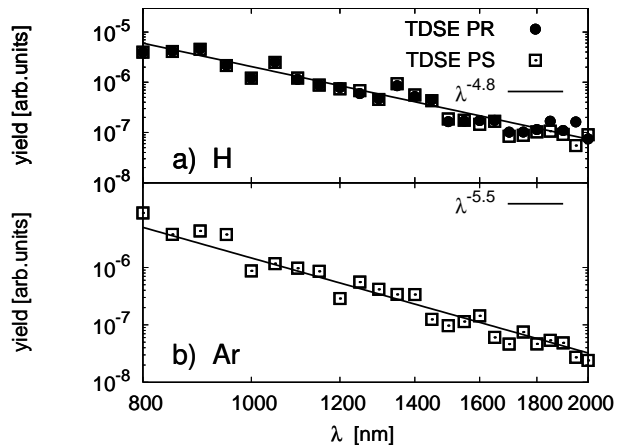,width=8.3cm,clip=} 
\caption{Integrated harmonic yield $\Delta I$ between 20 and 50 eV as a function of $\lambda$ calculated on a coarse mesh with $\Delta \lambda = 50$ nm. 
$\bullet$: Peaceman-Rachford method; $\Box$: pseudo-spectral method,  solid line: fit $\Delta I \propto \lambda^{-x}$: a) hydrogen, b) argon.}
\label{fi:global_scaling}
\end{figure}

While we present the results obtained from the direct Fourier transform $a(\omega)$ of the dipole acceleration, we have confirmed the application of a Welch or Bartlett window \cite{nr} in the transformation hardly affects the results except for a constant factor. 
The HHG yield (defined as radiated energy per unit time, \cite{Jackson}) integrated from 20 to 50 eV,
\begin{equation}
\Delta I  =  \frac{1}{3 c^3} \int_{20\,{\rm eV}}^{50\,{\rm eV}}  | a(\omega)|^2 d \omega %\frac{1}{T_p}
\label{eq:yield}
\end{equation}
calculated on a coarse mesh in $\lambda$ with a spacing of 50 nm (Fig.\ \ref{fi:global_scaling}) falls off with a power law, $\Delta I \propto \lambda^{-x}$ ($x\approx 4.8-5.5$) for H and Ar,
 in qualitative agreement with Ref.\ \cite{ar:tate_scaling07}.
The two alternative integration algorithms employed in this work agree well with each other. 
Small discrepancies near 2 $\mu$m are due to the difference in grid spacing and can be controlled by changes in the spacing near the origin.
A power law ($x \approx 5$) results from the combination of two effects: the spreading of the returning wavepacket would give $x=3$ \cite{ar:lewenstein94} for the overall yield. 
The increase of the cutoff $E_c \propto \lambda^2$ results for a fixed energy interval (see Eq.\ (\ref{eq:yield})) in an additional factor $\lambda^{-2}$.

A closer look at Fig.\ \ref{fi:global_scaling} reveals the remarkable feature that the harmonic yield does not vary smoothly with $\lambda$ as may have been anticipated in the previous work, but strongly fluctuates.
Slight change in fundamental wavelength may lead to variations of the yield by a factor of 2 to 6. 
Such rapid fluctuations imply that a reliable $\lambda$  dependence can only be established by employing a much finer $\lambda$ grid. 
Moreover, the notion of a simple power law scaling itself is called into question and can apply, if at all, only after averaging over fluctuations.
The fluctuations are not specific to hydrogen but appear for argon  (Fig.\ \ref{fi:global_scaling}(b)) as well.

%\section{Small-scale variations}

\begin{figure}
\epsfig{file=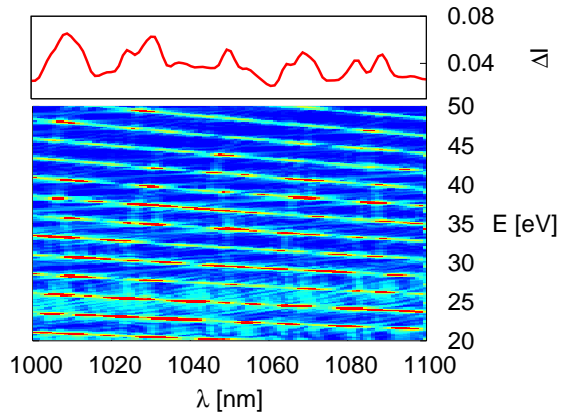,width=8.3cm,clip=} 
\caption{(Color online) Fluctuations of the harmonic yield $\Delta I$ as a function of the fundamental wavelength $\lambda$ and the photon energy $E$ for hydrogen.
%, quasi-periodic enhancements encircled.
An oscillatory behaviour of the yield as a function of $\lambda$ emerges after integrating over vertical strips (upper panel and Fig.\ \ref{fi:details_hydrogen1000}(a)). 
}
\label{fi:scaling3d}
\end{figure}

\begin{figure}
\epsfig{file=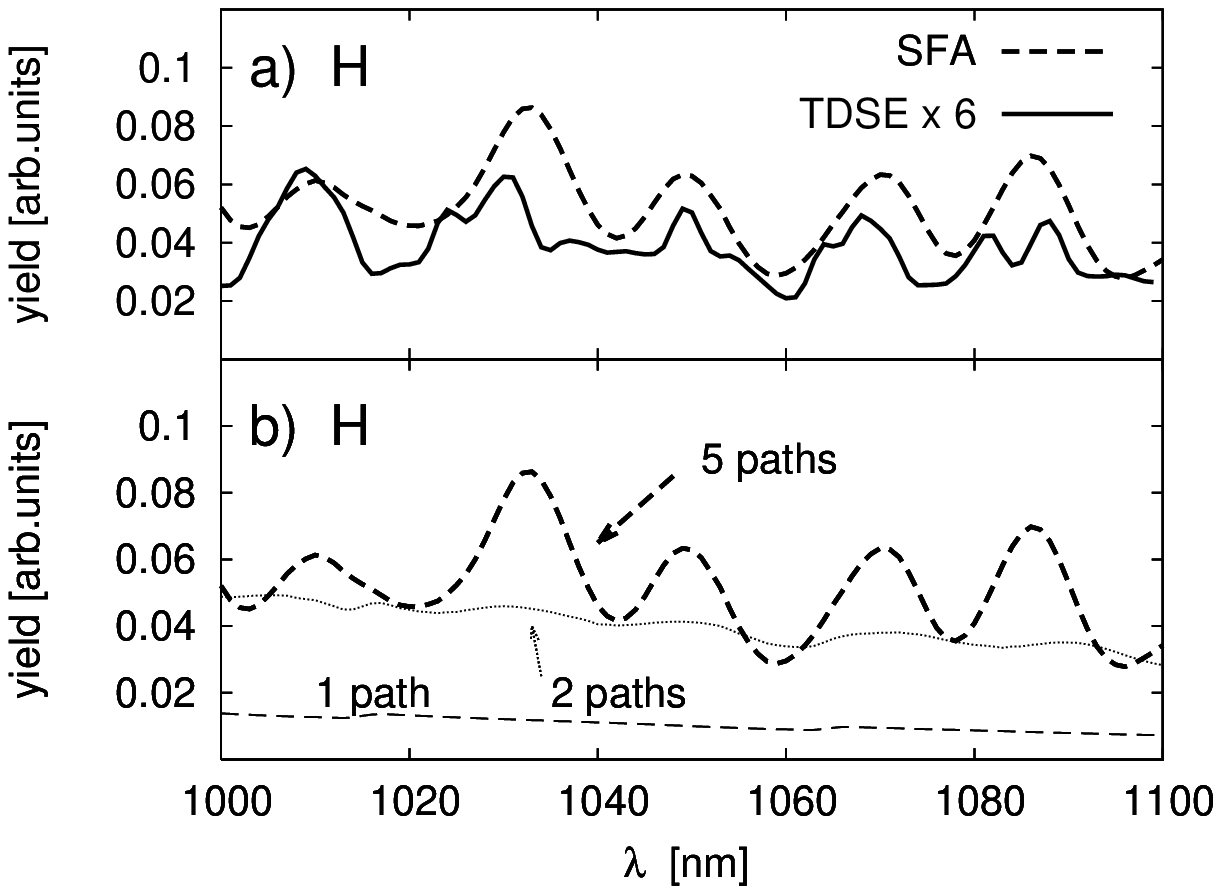,width=8.3cm,clip=}
\epsfig{file=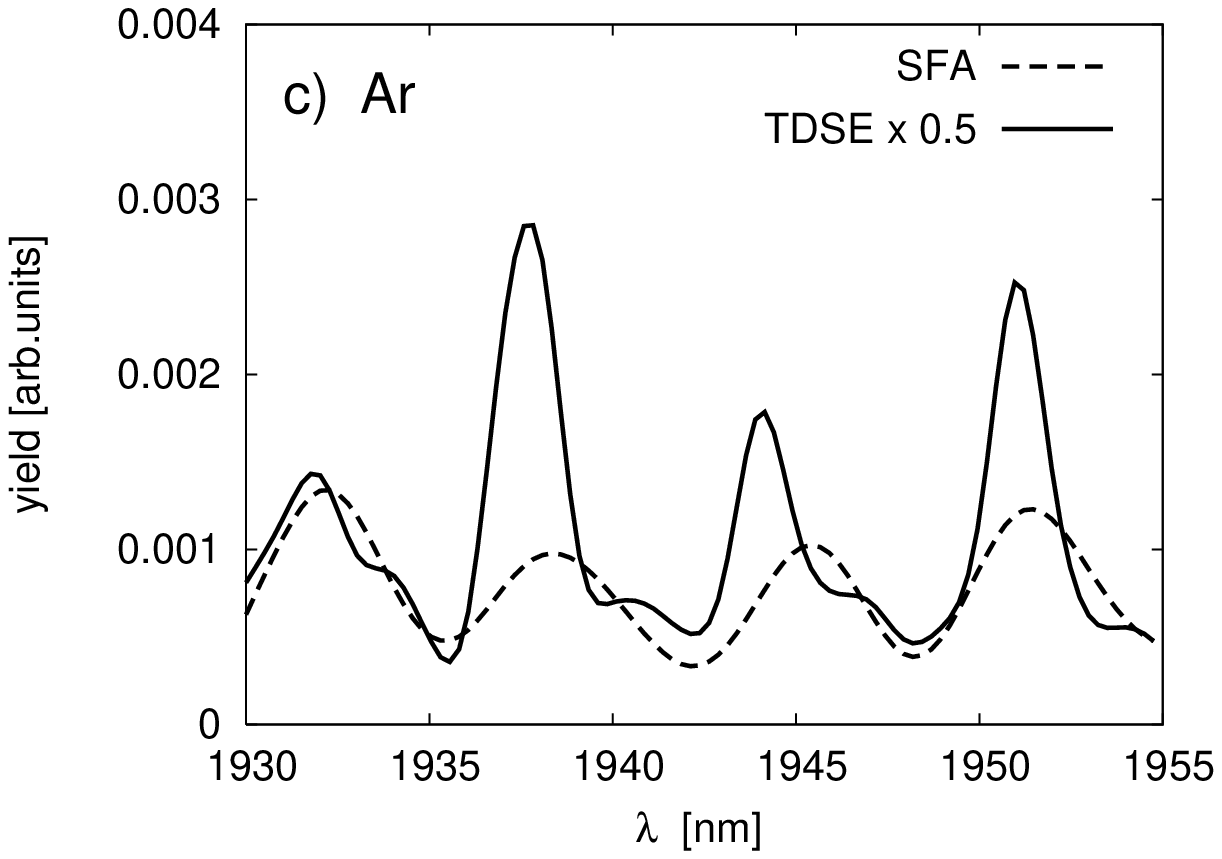,width=8.3cm,clip=}
\caption{
Variations in the integrated harmonic yield (20 to 50 eV) in a narrow range of $\lambda$.
a) comparison between the TDSE solution with the SFA including 5 quantum paths for hydrogen.
b) build-up of the interference pattern with increasing number of quantum trajectories within the SFA.
Thin dashed: 1 (short)  trajectory,
Thin dotted: 2 (short and long) trajectories,
Bold dashed: inclusion of the shortest 5 trajectories.
c) as a) but for argon.
\label{fi:details_hydrogen1000}}
\end{figure}

Figure \ref{fi:scaling3d} presents a two-dimensional zoom into the fine-scale variations calculated on a mesh with $\Delta \lambda=1$ nm, as a function of $\lambda$ and the photon energy $E$.
Fluctuations appear which form vertical ``ridges'' spanning several harmonic orders. 
Consequently, after integrating over vertical strips we arrive at a fluctuating function $\Delta I (\lambda)$  (upper panel).
%($E_1 \le E \le E_2$) 
Such variations in quantities averaged over a large energy interval are expected when few (quantum) orbits with a well-defined time structure significantly influence the spectrum (``periodic orbit spectra'') \cite{ar:periodic_orbit}.
A superposition of oscillations of different frequencies emerges (Fig.\ \ref{fi:details_hydrogen1000}), the dominant of which has a fluctuation scale $\delta \lambda$ of $\approx 20$ nm near a driver wavelength of $\lambda=1000$ nm, and  $\delta \lambda \approx 6$ nm near  $\lambda=2000$ nm.
Similar oscillation patterns can be observed for argon on Fig.\ \ref{fi:details_hydrogen1000}(c). 
Apparently, they are largely independent of the atomic species.
Such oscillations are obviously the result of interference effects.

Many features of HHG can be intuitively and even quantitatively explained in terms of quantum trajectories \cite{ar:lewenstein94,ar:salieres_science01} which represent the semiclassical three step model \cite{ar:corkum93}. 
The main contribution to the HHG spectrum
comes from those electronic quantum paths that correspond to classical returning trajectories ionized at a certain time $t_i$ and recombining with the parent ion at a later time $t_f$. 
In order to identify the origin of the interference structures we apply a semiclassical model based on the strong field approximation (SFA) \cite{ar:lewenstein94,ar:ivanov96}. 
In this model, the time-dependent dipole moment $d(t)$ can be expressed as \cite{ar:ivanov96} 
\begin{eqnarray}
d(t_f)= \sum_{P(t_i)} b_{\rm ion}(t_i) \cdot e^{ - i S_P(t_i,t_f) } \cdot c_{\rm rec}(t_f) + c.c.
\label{eq:SFA_dipole}
\end{eqnarray}
i.e.\ a sum over paths $P$ that start at the moment of tunnel ionization $t_i$ with amplitude $b_{\rm ion}(t_i)$, 
evolve in the laser field  - $e^{ - i S_P(t_i,t_f) }$ - and recombine upon rescattering at the core at time $t_f$ with the amplitude $c_{\rm rec}(t_f)$.
The sum over all possible electron trajectories recolliding at time $t_f$ may be large, but in practice is limited by wave-packet spreading.
We consider up to 16 possible ionization times $t_i$ for each individual $t_f$. 

Interference oscillations are controlled by the evolution phase, the semiclassical action of the path $P$, which reads:
\begin{eqnarray}
S_P(t_i,t_f) = \int_{t_i}^{t_f} \frac{(p + A(t'))^2}{2} dt' +g \cdot  |E_i| (t_f - t_i)
\label{eq:semiclassical_phase}
\end{eqnarray}
$|E_i|$ is the ionization potential (binding energy) of the atom and $A(t)$ the laser vector potential defined by $A(t)=-\int_t^T F(t') dt'$.
$p$ is the classical momentum of the returning trajectory.
$g=1.3$ is a correction factor that accounts for the dependence on the tunneling time, modifying the energy of the recolliding photons \cite{ar:lewenstein94,ar:vlad_armin03}.
When including up to five returning paths, the semiclassical calculation can reproduce the modulation depth, 
modulation frequency, and the approximate phase of the $\lambda$ oscillations reasonably well, thus unambiguously establishing the quantum path interference as the origin of the fluctuations (Fig.\ \ref{fi:details_hydrogen1000}(b)).
Consideration of additional trajectories leads only to minor modifications with no qualitative differences.
Setting $g=1.0$ yields only a phase shift of the predicted oscillations while retaining the modulation frequency and depth.

Very recently, first experimental evidence for the interference between the short and long paths was presented \cite{ar:zair_icpeac07} in the \textit{intensity dependence} of the HHG yield.
Remarkably, for the present $\lambda$ dependence the frequently discussed short and long trajectories (dotted curve in Fig.\ \ref{fi:details_hydrogen1000}(b)) are insufficient to account for the interference oscillations. 
Convergence is approximately reached only when at least five trajectories are included.
Moreover, the presence of the oscillations is independent of the particular choice of the envelope function $f(t)$. 
Note that the Fourier broadening of the few-cycle driving field exceeds, on a wavelength scale, the period $\delta \lambda$ of the modulation.
This somewhat surprising finding is a direct consequence of the quantum path interference. 
As long as the few-cycle pulse permits the generation of a set of a few quantum paths in subsequent half-cycles, the overall temporal characteristics of the driver pulse is of minor importance, 
though the latter will influence the detailed shape of the interference pattern.
Preliminary calculations for pulse propagation in one dimension, accounting for the geometric Guoy phase, show that inteference oscillations persist in loose focus geometry.

\begin{figure}
\epsfig{file=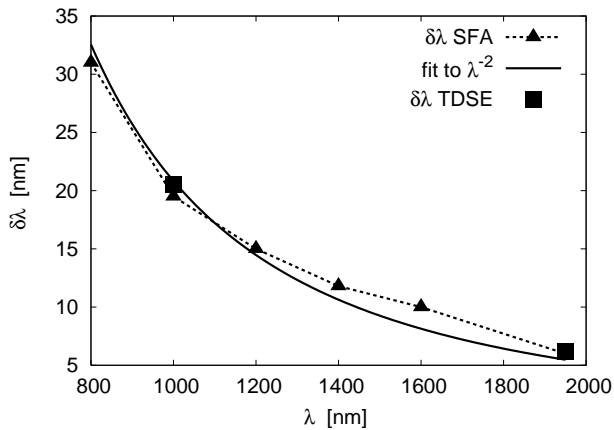,width=8.3cm,clip=}
\caption{
Variation of the modulation period $\delta \lambda$ with the fundamental wavelength $\lambda$ for hydrogen.
$\blacksquare$: TDSE, $\blacktriangle$: SFA.
Solid line: $ \lambda^{-2}$ scaling (see Eq.\ (\ref{eq:formula_deltalambda1})).
\label{fi:scaling_deltalambda}}
\end{figure}

The principal modulation length of the harmonic yield $\delta \lambda$ is a function of the wavelength $\lambda$ itself.
It approaches $\approx 6$ nm near a wavelength of 2 $\mu$m (Fig.\ \ref{fi:scaling_deltalambda}).
For a simple estimate  for the scaling of $\delta \lambda$ with $\lambda$ we note that the
semiclassical action in Eq.\ (\ref{eq:semiclassical_phase}) has its largest contribution from the $A(t)^2$ term in the strong field case.
Hence, the phase difference between the shortest and longer tajectories due to the semiclassical action can be approximated by $\bar S_P \approx U_p \cdot \tau_f$ where $ \tau_f=t_f-t_i$ is the flight time of the electron trajectory \cite{ar:salieres_science01}. 
$\bar S_P$ scales approximately as $\lambda^{3}$.
The period of the modulation corresponds to a phase change of $\bar S_P$ by $2\pi$. Accordingly, 
\begin{eqnarray}
\label{eq:variation_phase}
2\pi = \delta \bar S_P = \frac{dS}{d\lambda} \delta \lambda \\
{\rm or}~~~~ \delta \lambda  \propto \lambda ^{-2}
\label{eq:formula_deltalambda1}
\end{eqnarray}
%\begin{equation}
%\label{eq:variation_phase}
%2\pi = \delta \bar S_P = \frac{dS}{d\lambda} \delta \lambda 
%\end{equation}
%or
%\begin{equation}
%\delta \lambda  \propto \lambda ^{-2}
%\label{eq:formula_deltalambda1}
%\end{equation}
This estimate can be improved when using the full expression for $S_P$; % (Eq.\ (\ref{eq:semiclassical_phase})).
nevertheless even Eq.\ (\ref{eq:formula_deltalambda1}) predicts the $\lambda$ dependence of the modulation length remarkably well (see Fig.\ \ref{fi:scaling_deltalambda}).

%\section{Summary}
In conclusion, we have found that the fundamental wavelength dependence of HHG in the single-atom response features surprisingly strong oscillations on fine wavelength scales with modulation periods as small as 6 nm in the mid-infrared regime near $\lambda = 2\, \mu$m. 
These oscillations have been established for both hydrogen and rare gas targets (as an example argon is shown in this work) using two complementary integration algorithms for the TDSE.
Thus, even a slight change in fundamental wavelength leads to variations in the HHG yield by a factor of up to $\approx 6$. 
According to our semiclassical analysis based on the SFA, this unexpectedly rapid variation on a fine scale is the consequence of the interference of different rescattering trajectories. 
We have to take account of up to five returns to reproduce the results of the quantum simulations. 
This confirms the significance of higher-order returns of the electron wavepacket \cite{ar:tate_scaling07}. 
On a large $\lambda$ scale, apart from the rapid oscillation, our TDSE results show that the HHG yield at constant intensity decreases as $\lambda^{-x}$ with $x \approx 5$ for H and Ar.
This dependence is different from the generally accepted $\lambda^{-3}$ scaling, but is close to that reported in Ref.\ \cite{ar:tate_scaling07}.

\begin{acknowledgments}
The work was supported by 
the Austrian ``Fonds zur F\"orderung der wissenschaftlichen Forschung'', under grant no. FWF-SFB016 ``ADLIS''.
K.\ S.\ also aknowledges support by the IMPRS-APS program of the MPQ (Germany). K.L.I. gratefully acknowledges financial support by the Precursory Research for Embryonic Science and Technology (PRESTO) program of the Japan Science and Technology Agency (JST) and by the Ministry of Education, Culture, Sports, Science, and Technology of Japan, Grant No. 19686006.
\end{acknowledgments}

\end{document}